\documentclass[pra,aps,twocolumn,showpacs,superscriptaddress,amsmath,amssymb]{revtex4-1}
\usepackage{color}
\usepackage[T1]{fontenc}
\usepackage{amsmath,graphicx}

\begin{document}
\def\tr{\rm{Tr}}
\def\la{{\langle}}
\def\ra{{\rangle}}
\def\a{{\alpha}}
\def\e{\epsilon}
\def\q{\quad}
\def\w{\tilde{W}}
\def\t{\tilde{t}}
\def\a{\hat{A}}
\def\h{\hat{H}}
\def\E{\mathcal{E}}
\def\p{\hat{P}}
\def\u{\hat{U}}

\title{Schroedinger current for discontinuous states from the first passage time decomposition}
%
%
\author {D. Sokolovski}
\affiliation{Departamento de Qu\'imica-F\'isica, Universidad del Pa\' is Vasco, UPV/EHU, Leioa, Spain}
\affiliation{IKERBASQUE, Basque Foundation for Science, E-48011 Bilbao, Spain}

\date{\today}
\begin{abstract}
We revisit the problem of calculating the probability current for discontinuous states, such that may 
arise in atom trapping or as a result of projective measurements. In the first passage time representation, the problem reduces to evaluation of a localised wave originating from the discontinuity, whose interference with the initial state determines the transfer of probability. 
Depending on the type of discontinuity, the current behaves as $t^{1/2}$, $t^{3/2}$ or $const(t)$.
Our approach generalises earlier work on this subject.

\end{abstract}

%
%
\maketitle
%
%
%
%
%
%
%
\section{Introduction}
Quantum states whose wavefunctions experience, at certain points, discontinuity in the value or the first derivative may arise in such applications as atom trapping by means of strong laser beams \cite{RAIZ1}, or quantum measurements aimed at detecting a particle in a given region of space \cite{SCHUSS}.
One may wish to estimate the rate at which the probability is transferred across a discontinuity, for example, because it determines the presence and the type of the Zeno effect \cite{SCHUSS}-\cite{ZDS}.
This rate cannot, however, be obtained by means of the conventional current operator owing to the singular behaviour of the kinetic energy. In Ref.\cite{SCHUSS} the authors evaluated the current at the edge of a truncated state by integrating the Feynman propagator over the region of support of the original wavefunction. The purpose of this paper is to generalise their results to the case of an arbitrary discontinuity which may occur in the initial state. For this we will employ the quantum first passage time  (FPT) representation  \cite{FP1}-\cite{FP4}, 
leading to the physically appealing picture in which a localised wave is emitted from the point where the the wavefunction or its first derivative suffers a  jump, and whose interference with the rest of the state determines the probability current at the discontinuity. The rest of the paper is organised as follows: in Sect.II we introduce the  FPT and in Sect. III we define the emitted wave, $\delta\Psi(x,t)$, whose purpose is to repair the initial discontinuity at finite times. In Section IV we use the conservation of probability to 
derive some useful properties of $\delta\Psi(x,t)$. Section V gives the short time limit of the probability current and specialises to a number of particular cases. Section V contains our conclusions.
\section{Evolution of the wavefunction in the first passage time representation}
For a particle of  mass $M$ in one dimension, we wish to evaluate the probability current for a complex valued wavefunction 
$\Psi(x)=\Psi_1(x)+i\Psi_2(x)$, 
at a point which we choose to be the origin, $x=0$.
The particle moves in a potential $V(x)$ so that the Hamiltonian takes the form (we use $\hbar=1$)
\begin{eqnarray}\label{1a}
\h=-\partial_x^2/2M+V(x),
\end{eqnarray}
and without loss of generality we choose the $\Psi(x)$ to vanish at some  $a < 0$ and $b>0$ .
The task is not trivial since we will assume that both the wavefunction and its first derivative may be discontinuous at $x=0$, i.e.,  ($f(\pm0)\equiv lim{\epsilon \to 0} f(\pm \epsilon)$)
\begin{eqnarray}\label{2}
\Psi(-0)\ne \Psi(+0), 
\q \Psi'(-0)\ne \Psi'(+0),
\end{eqnarray}
where
the prime denotes differentiation with respect to $x$.
Thus, we have 
\begin{eqnarray}\label{3}
\Psi(x)=\Psi^L(x)\theta_L(x)+\Psi^R(x)\theta_R(x),
\end{eqnarray}
where $\theta^L(x)=1$ for $x\le0$ and $0$ otherwise, and $\theta^R(x)=1-\theta^L(x)$.
To analyse the development of, say, $ |\Psi^L\ra \equiv \int_a^0dx\Psi^L(x)|x\ra$ we invoke the first crossing time decomposition of a general evolution operator discussed in \cite{FP1}-\cite{FP4}
\begin{eqnarray}\label{4}
\exp(-i\h t)|\Phi\ra=\p\exp(-i\p\h\p t)\p|\Phi\ra+\q\q\q\q
\\\ \nonumber
-i\int_0^tdt_1 \exp[-i\h(t-t_1)][\h,\p]\p\exp(-i\p\h\p t_1)\p|\Phi\ra
\end{eqnarray}
which describes the time evolution, with a Hamiltonian $\h$, of a state initially localised within
a subspace $\mathcal{H_P}$ onto which projects the projector $\p$, $\p|\Phi\ra=|\Phi\ra$.
Equation (\ref{4}) has the standard interpretation \cite{FP4}: the system moves within $\mathcal{H_P}$ until a time $t_1$ when it leaves the subspace for the first time, after which it may or may not reenter it. The first term in the r.h.s. of Eq. (\ref{4}) corresponds to the scenario in which the system has not left $\mathcal{H_P}$ by the time $t$.
\newline
The choice of the projector $\p$ appropriate for our purpose requires some care. We begin with the real part of $\Psi^L(x)$, $\Psi^L_1(x)$, define $\alpha_1$ so that
 $\Psi^L_1(0)+\alpha_1\Psi^{L'}_1(0)=0$, and then choose
$\p_{\alpha_1}=\sum_n |\phi^L_n\ra \la\phi^L_n|$
with $\h|\phi^L_n\ra=E^L_n|\phi^L_n\ra$, subject to the boundary conditions
$\phi^L_n(a)=0,\q \phi^L_n(0)+\alpha_1\phi^{L'}_n(0)=0$,
and the additional requirement that
$\phi^L_n(x)\equiv0$, for $0< x\le b$.
Since the eigenfunctions of a real Sturm-Liouville problem form a complete orthonormal basis, the operator $\p_{\alpha_1}$ is indeed the projector on a functional space containing our  state of interest, $\Psi^L_1(x)$. With this, the reduced evolution operator in Eq.(\ref{4}) takes the form
 \begin{eqnarray}\label{7a}
\p_{\alpha_1}\exp(-i\p_{\alpha_1}\h\p_{\alpha_1} t)\p_{\alpha_1}=
\\ \nonumber \sum_n |\phi^L_n\ra \exp(-iE_n^Lt)\la\phi^L_n|\equiv \u_{\alpha_1}(t).
\end{eqnarray}
The commutator $[\h,\p_{\alpha_1}]$ is independent of $\alpha_1$, as shown in the Appendix A, namely
$\la f|[\h,\p_{\alpha_1}]|g\ra =-[f^{*'}(0)g(0)-f^*(0)g'(0)]/2M$
  which provides the last ingredient required in Eq.(\ref{4}).
Repeating the above steps for imaginary parts of the left wavefunction, 
 $\Psi^L_2(x)$, introduces the parameter $\alpha_2$, 
$\Psi^L_2(0)+\alpha_2\Psi^{L'}_2(0)=0$.
We then have an exact result 
   \begin{eqnarray}\label{10} \nonumber
\la x|\exp(-i\h t)|\Psi^L\ra = \varphi_1(x,t|\alpha_1)+i\varphi_2(x,t|\alpha_2)
+\frac{i}{2M}\times \q\q\\ \nonumber\int_0^t dt_1
\{\partial_{x'}G'(x,x'=0,t-t_1)[ \varphi_1(0,t|\alpha_1)+i\varphi_2(0,t|\alpha_2)]\q\q
\\ 
-G(x,0,t-t_1)[ \varphi'_1(0,t|\alpha_1)+i\varphi'_2(0,t|\alpha_2)]\}\q\q
\end{eqnarray}
where $\varphi_{1,2}(x,t|\alpha_{1,2})\equiv \la x|\u_{\alpha_{1,2}}(t)|\Psi^L_{1,2}\ra$ are the results of evolving the real and imaginary parts of the wavefunction on an interval $[a,0]$ with 
the Hermitian Hamiltonians corresponding to (possibly) non-physical boundary conditions 
at $x=0$,
 $\varphi_{1,2}(x,t|\alpha_{1,2})=-\alpha_{1,2}\varphi'_{1,2}(x,t|\alpha_{1,2})$.
 and  $G(x,x',t)\equiv \la x|\exp(-i\h t)|x'\ra$,
\section{ Smoothing of the discontinuity}
Evaluation of  the current at $t=0$ only requires the short time limit of the Eq.(\ref{10}), 
which allows us to put 
$\u_{\alpha_1}(t)\approx \u_{\alpha_2}(t)\approx 1$,
in the integral
and
replace the terms in the square brackets by
$\Psi(-0)$ and $\Psi'(-0)$, respectively. Further, for a small $t$, the full propagator 
may be replaced by the free-particle one \cite{FEYN},\cite{MILL} $G(x,x',t) \approx G_0(x-x',t) \equiv\q\q\q\q\q\q\q\q\q\q\q\
 \sqrt{M/2\pi it}\exp[iM(x-x')^2/2t]$, with which Eq.(\ref{10}) reduces to [$\Psi^L(x,t)\equiv \la x|\exp(-i\h t)|\Psi^L\ra$]
    \begin{eqnarray}\label{b3}
\Psi^L(x,t)\approx \theta_L(x)[\Psi^L(x) -i\la x|\h|\Psi^L\ra t]
\\ \nonumber
+\Psi(-0)\delta \Psi'(x,t)+\Psi'(-0)\delta \Psi(x,t).
\end{eqnarray}
We note that $\delta \Psi(x,t)$ is superposition of all waves emitted by a constant point source 
 at the origin for all $0\le t_1\le t$,
\begin{eqnarray}\label{b4}
\nonumber
\delta \Psi(x,t) \equiv -\frac{i}{2M}\int_0^t G_0(x,t_1)dt_1=\q\q\q\q\q\q
\\ 
-\sqrt{\frac{it}{2M\pi}}\exp(iMx^2/2t)+
\frac{|x|}{2}
 { erfc}(\sqrt{Mx^2/2it}),\q
\end{eqnarray}
where $erfc(z)=(2/\sqrt{\pi}) \int_z^{\infty}\exp(-z^2)dz$ is the complimentary error function \cite{AST}.
It is readily seen that  $\delta \Psi(x,t)$, continuous with discontinuous first derivative,  
\begin{eqnarray}\label{b2}
\delta \Psi(0)=-\sqrt{\frac{it}{2M\pi}},\q
\delta \Psi'(+0)=-\delta \Psi'(-0)=1/2,\q
\end{eqnarray}
serves to repair  initial discontinuities in the value 
and the first derivative
 of  $\Psi^L(x)\theta_L(x)$ at $x=0$, as illustrated in Fig.1. Indeed, from Eqs.(\ref{b3}) and (\ref{b4}) for $t>0$ we have
  \begin{figure}[h]
\includegraphics[width=7cm, angle=0]{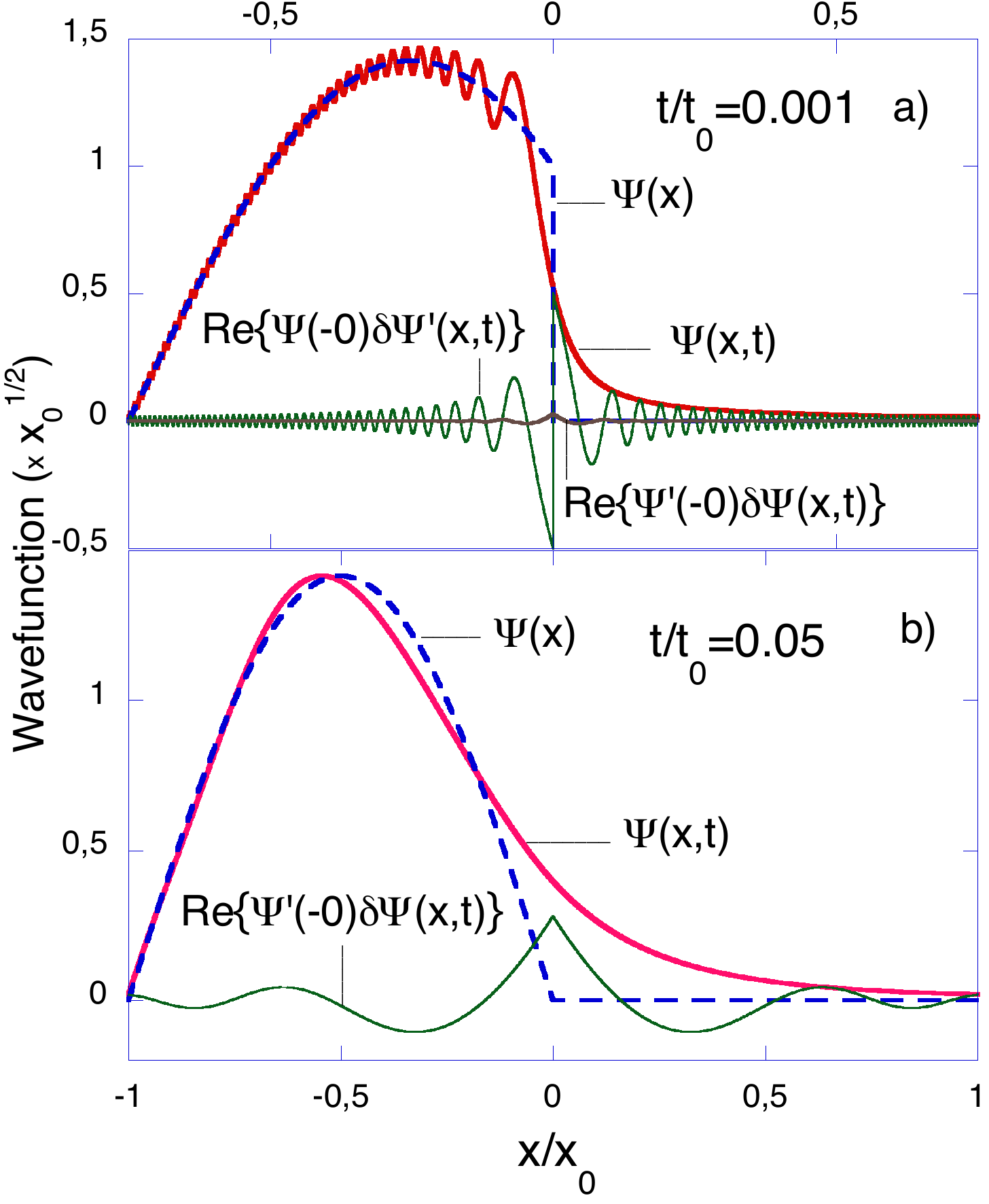}
\caption{(colour online) Short time evolution of the ground state of an infinite well
of a length $x_0$ (thick solid): a) truncated at three quarters of its length (thick dashed),
for $t/t_0=0.001$, $t_0\equiv Mx_0^2$; b)
with the right wall of the well removed (thick dashed), for $t/t_0=0.05$.
Also shown are various terms in Eq.(\ref{b3}).} 
\label{FIG3}
\end{figure}
    \begin{eqnarray}\label{b5}\nonumber
\Psi^L(+0,t) = \Psi^L(-0,t)=\Psi(-0)/2+\delta \Psi(0,t)\Psi'(-0)+O(t) \q\q\q \\ \nonumber
\Psi^{L'}(+0,t)=\Psi^{L'}(-0,t)=\Psi'(-0)/2+\delta \Psi''(0)\Psi(-0)+O(t)\q\q\q
\end{eqnarray}
We note also that $\delta \Psi(0,t)$ is highly oscillatory and rapidly decays away from the origin,
\begin{eqnarray}\label{b6}
lim_{|x| \to \infty }\delta \Psi(x,t)=\frac{1}{\sqrt{2\pi i}x^2}\left ( \frac{t}{M}\right) ^{3/2}\exp(iMx^2/2t).\q\q
\end{eqnarray}
Finally,  changing $x\to -x$, we can repeat the calculation for $\Psi^R$. Adding the two contributions then yields
\begin{eqnarray}\label{b7}
\Psi(x,t) \approx \Psi(x)
 \nonumber
-i[\theta^L(x)\la x|\h|\Psi^L\ra 
+\theta^R(x)\la x|\h|\Psi^R\ra] t\\
-[\delta \Psi'(x,t)\Delta \Psi 
+\delta \Psi(x,t)\Delta \Psi'],\q\q\q\q\q\q\q\q\q
\end{eqnarray}
\begin{eqnarray}\label{b7a}\nonumber
\Delta \Psi \equiv \Psi(+0)-\Psi(-0), \q\Delta \Psi' \equiv \Psi'(+0)-\Psi'(-0)\q 
\end{eqnarray}
where the second term is responsible for the Schroedinger current in a continuous state (see below), while the last term describes an additional wave emitted at the point of discontinuity. 
\section{Conservation of probability}
The evolution of the wavefunction is unitary, which implies
$\int_a^b |\Psi(x,t)|^2dx = \int_a^0|\Psi^L(x)|^2dx+\int_0^b|\Psi^L(x)|^2dx$.
With a discontinuity characterised by $\Psi(\pm0)$ and  $\Psi'(\pm0)$, there are (discounting an overall phase) seven independent real parameters in Eq.(\ref{b7}).
Inserting Eq.(\ref{b7}) into the above yields, therefore, seven conditions involving integrals 
of $\delta \Psi(x,t)$ and $\delta \Psi'(x,t)$, two of which we will consider next.
 Since $\delta \Psi(x,t)$ is highly oscillatory except in the vicinity of $x=0$ [cf. Eq.(\ref{b6})], we may expand $\Psi^{L,R}(x) \approx \Psi(\mp0)+\Psi'(\mp0)x$ and extend the integration to $\pm \infty$. 
Choosing  $\Psi^R(x)\equiv 0$ and $\Psi'(-0)=0$ Eq.(\ref{b7}) then yields
\begin{eqnarray}\label{c2}
\int_0^{\infty} |\delta\Psi'(x,t)|^2dx =-Re \int_{-\infty}^0\delta\Psi'(x)dx
=\frac{t^{1/2}}{2\sqrt{M\pi}}\q\q.
\end{eqnarray}
\newline
Similarly, for $\Psi^R(x)\equiv 0$ and  $\Psi(x)$ vanishing at the origin, $\Psi(-0)=0$,
we have
\begin{eqnarray}\label{c3}
\int_0^{\infty} |\delta\Psi(x,t)|^2dx = Re \int_0^{\infty} x\delta\Psi(x)dx
=\frac{t^{3/2}}{6\sqrt{M^3\pi}},\q\q
\end{eqnarray}
which can, if one wishes, be verified directly (see Appendix C).
In the next Section we will also need the relations
\begin{eqnarray}\label{c4}
\int_0^{\infty} \delta\Psi(x,t)dx=
-\int_0^{\infty}x \delta\Psi'(x,t)dx = \frac{t}{4Mi},
\end{eqnarray}
which cannot be deduced from conservation of probability, but are readily obtained by integrating Eq.(\ref{b4}).
\section{The Schroedinger current}
The probabilty current at the discontinuity, $x=0$, is defined as
\begin{eqnarray}\label{d1}
J(t)=\partial_t \int_0^\infty |\Psi(x,t)|^2dx\equiv\partial_t P^R(t)
\end{eqnarray}
with the probability $P^R(t)$ determined by the interference between various terms in Eq.(\ref{b7}).
Retaining the principal contributions, we have 
\begin{eqnarray}\label{d2}\nonumber
P^R(t)\approx\int_0^b|\Psi^R(x)|^2dx +\q\q\q\q\q\q\q\q\q\q\q\\ \nonumber
\{[|\Psi(-0)|^2-|\Psi(+0)]|^2Re\int_0^\infty \delta \Psi'(x,t)dx-\q\q\q\\
2|\Psi(-0)||\Psi(+0)|\sin\Delta\phi_1 Im \int_0^\infty \delta \Psi'(x,t)dx\} \q\q\q\\ \nonumber
+\{\frac{t}{M}Im[\Psi^*(+0)\Psi'(+0)]+\q\q\q\q\q\q\q\q\q\q\q\\ \nonumber
2Re[(\Psi^{*'}(+0)\Delta \Psi-\Psi^*(+0)\Delta \Psi') \int_0^\infty \delta \Psi(x,t)dx+
\end{eqnarray}
\begin{eqnarray}\nonumber
\Delta \Psi^{*'}\Delta \Psi
\int_0^\infty\delta \Psi^*(x,t)\delta \Psi'(x,t)]dx\}\q\q\q\q\q\q\q\\ \nonumber
+\{[|\Psi'(-0)|^2-|\Psi'(+0)|^2]Re\int_0^\infty x \delta \Psi(x,t)dx\q\q\q\\ \nonumber-
2|\Psi'(-0)||\Psi'(+0)|\sin\Delta\phi_2 Im \int_0^\infty x\delta \Psi x,t)dx\}\q\q \\ \nonumber
\end{eqnarray}
where $\Delta \phi_1\equiv \arg[\Psi(+0)]- \arg[\Psi(-0)]$, $\Delta \phi_2\equiv \arg[\Psi'(+0)]- \arg[\Psi'(-0)]$. Above we have used Eqs.(\ref{c2}), (\ref{c3}) and (\ref{c4}),
which indicate that in Eq.(\ref{d2}) the terms in the first, second and third curly brackets 
are proportional to $t^{1/2}$, $t$ and $t^{3/2}$, respectively \cite{FOOTOM}. Thus,
 to the leading order in $t$,we have:
\newline
{\it A. Continuous state.} If no discontinuity is present, 
$\Delta \Psi'=\Delta \Psi=0$, Eq.(\ref{d2}) reduces to the first term in the second 
curly bracket and, as $t\to 0$, the current $J(t)$ takes the usual form,
\begin{eqnarray}\label{d3}
J(t)\approx \frac{i}{2M}[\Psi^{*'}(0)\Psi(0)-\Psi^*(0)\Psi'(0)]\sim const(t)\q\q
\end{eqnarray}
\newline
{\it B. Discontinuity in the value of the function.} For $\Delta \Psi\ne 0$, 
regardless of the values of $\Delta \Psi'$, $\Psi(+0)$ and $\Psi'(+0)$,
Eq.(\ref{d2}) is dominated by the terms in the first curly bracket, and the 
current behaves as $t^{-1/2}$ (see also Ref.\cite{SCHUSS}),
\begin{eqnarray}\label{d4}
J(t)\approx \{[|\Psi(-0)|^2-|\Psi(+0)]|^2\q\q\q\q\q\q\\ \nonumber
-2|\Psi(-0)||\Psi(+0)|\sin\Delta\phi_1\}\frac{1}{4\sqrt{M\pi t}} 
\sim t^{-1/2}.
\end{eqnarray}
Note that if the modulus of the initial wavefunction is continuous, 
the current is proportional to the sine of the jump in the phase, $\Delta \phi_1$.\newline
{\it C. Discontinuity in the first derivative, and $\Psi(0)\ne 0$.} 
For $\Delta \Psi =0$, $\Delta \Psi'\ne 0$ and $\Psi^R(x)\ne 0$
the leading contribution to the current comes from the first two terms in the second curly bracket of 
Eq.(\ref{d2}), and using Eq.(\ref{c4}) we have
\begin{eqnarray}\label{d5}
J(t)\approx -Im\{\Psi^*(0)[\Psi'(+0)+\Psi'(-0)]/2\}\sim const (t),\q\q
\end{eqnarray}
which reduces to Eq.(\ref{d3}) if $\Psi'(+0)=\Psi'(-0)$.
\newline
{\it D. Discontinuity in the first derivative, and $\Psi(0)= 0$.} 
For $\Delta \Psi =0$, $\Delta \Psi'\ne 0$ and $\Psi^R(0)=0$
from Eqs.(\ref{d3}) and (\ref{ac2}) we have
\begin{eqnarray}\label{d5}
J(t)\approx \{[|\Psi'(-0)|^2-|\Psi'(+0)|^2]/4+\q\q\q\q\\ \nonumber
|\Psi'(-0)||\Psi'(+0)|\sin\Delta\phi_2/2\} \sqrt{\frac{t}{\pi M^3}}\sim t^{1/2}.
\end{eqnarray}
Note that if the right hand space is empty, $\Psi^R(x)\equiv 0$, Eq.(\ref{d5}) reduces to the result of 
Ref.\cite{SCHUSS}.

\section{Conclusions and discussion}
In summary, 
a discontinuity present in a wavefunction $\Psi(x)$ at $t=0$, can be seen, for $t>0$, as a source of an additional wave propagating on both sides of the point where $\Psi(x)$ and/or $\Psi'(x)$ experience jumps. At short times, the wave consists of the amplitude emitted by a constant point source, $\delta\Psi(x,t)$ in Eq, (\ref{b4}), and its spacial derivative, weighed by the jumps in the values of $\Psi'(x)$ and the $\Psi(x)$, respectively. These additional terms serve to repair initial discontinuities, and their interference with the rest of the wavefunction determines both direction and rate of the probability transfer. If the value of wavefunction is discontinuous, the probability is always transferred at a rate  $\sim \sqrt{t}$, and the current behaves as $t^{-1/2}$. In the special case where the modulus of $\Psi(x)$ is continuous, the current is proportional the the sine of the jump in its phase. As suggested in \cite{SCHUSS} such states can be obtained as a result of  non-detection of atom, which eliminates, partially or completely, its wavefunction in a specified region of space. For a discontinuity in the first derivative at a point where $\Psi(x)$ does not vanish we recover the constant current given by the usual expression (\ref{d3})  with $\Psi'(x)$ replaced by the mean $[\Psi'(-0)+\Psi'(+0)]/2$ [cf. Eq,(\ref{d5})].
Such states would arise after an instantaneous collapse of a narrow ($\delta$-) barrier of a finite transparency. Finally, for $\Psi(0)=0$, e.g., in a state obtained by switching off a laser induced atomic trap \cite{RAIZ1},\cite{ZDS},  the current increases as $t^{1/2}$ and contains a term proportional to the sine of the jump in the phase of $\Psi'(x)$.

This work was supported by the Basque Government grant IT472
and MICINN (Ministerio de Ciencia e Innovaci—n) grant FIS2009-12773-C02-01.
\section{Appendix A. Commutator of $P_{\alpha}$ with the Hamiltonian.}
We need to evaluate matrix elements of the commutator 
$\la f|[\h,P_{\alpha}]|g\ra=
\sum_n\{\la f|\h |\phi_n\ra\la\phi_n|g\ra-
 \la f|\phi_n\ra \la\phi_n|\h |g \ra\}$
 Differentiation and integration by parts yields (assuming $f(a)=g(a)=0$),
\begin{eqnarray}
\label{ap1}
\int_a^bf^*(x)[\phi_n(x)\theta_L(x)]''dx =\q\q\q\q\q\q\q\q\q\\ \nonumber
\int_a^0f^*(x)\phi''_n(x)dx+[f^{*'}(0)+f^*(0)/\alpha]\phi_n(0)
\end{eqnarray}
and
\begin{eqnarray}\label{ap3}
\int_a^b[\phi_n(x)\theta_L(x)]g''(x)dx =\q\q\q\q\q\q\q\q\q
 \\ \nonumber
=
\int_a^0g(x)\phi''_n(x)dx+[g'(0)+g(0)/\alpha]\phi_n(0).
\end{eqnarray}
With $-\phi_n''(x)/2M+V(x)\phi_n(x)=E^L_n\phi_n(x)$ and 
$\sum_n\phi_n(0)\phi_n(x)=\delta(x)$, we find that the terms
$\sum_n\la f|\phi_n\ra E_n \la \phi_n g\ra$
and
 $f^*(0)g(0)/\alpha$ cancel
and the commutator
$\la f|[\h,P_{\alpha}]|g\ra=-[f^{*'}(0)g(0)-f^*(0)g'(0)]/2M$
is independent of the choice of $\alpha$.
\newline
\section{Appendix B. An alternative derivation of Eq.(\ref{b7}).}
Writing $\Psi(x,t)=\Psi(x)+\Phi(x,t)$, and solving the Schroedinger equation for $\Phi$ 
using the Green's function technique yields
\begin{eqnarray}\label{ab1}
\Phi(x,t)=\Psi(x)-
i\int_0^\infty dt' \theta(t-t')\times \\ \nonumber\int_a^b dx' \la x |\exp[-i\h( t-t')]|x'\ra\la x'|\h|\Psi\ra  \q\q
\end{eqnarray}
Differentiating Eq.(\ref{3}) we have
\begin{eqnarray}\label{ab2} \nonumber
\la x|\h|\Psi^{L,R}\ra =\theta_{L,R}(x) [-\frac{1}{2M}\Psi^{L,R''}(x)+V(x)\Psi^{L,R}(x)]\q\\
\pm\frac{1}{2M}[2\delta(x)\Psi^{L,R'}(x)+\delta'(x)\Psi^{L,R}(x)]\q\q\q\q
\end{eqnarray}
Inserting, Eqs.(\ref{ab2}) into (\ref{ab1}) and sending $t\to 0$ we have 
Eq.(\ref{b7}).
\section{Appendix C. Verification of Eq.(\ref{c3}).}
Inserting Eq.(\ref{b4}) into (\ref{c3}) and performing the Gaussian integration over $x$
yields
\begin{eqnarray}\label{ac1}
I_1(t)\equiv \int_0^{\infty} |\delta \Psi(x,t)|^2dx=\q\q\q\q\q\q\\ \nonumber
(1/8\sqrt{\pi M^3})\int_0^tdt_1\int_0^{t_1}dt_2(t_1-t_2)^{-1/2}=t^{3/2}/6\sqrt{\pi M^3}
\end{eqnarray}
On the other hand, $xG_0(x,t)=(t/iM)G'(x,t)$ so that
\begin{eqnarray}\label{ac2}
I_2(t)\equiv \int_0^{\infty} x\delta\Psi(x,t)dx=(1/2M^2)\int_0^t dt' t'G_0(0,t')\q\q\\ \nonumber
=t^{3/2}/3\sqrt{2\pi i M^3}
\end{eqnarray}
 and $I_1=Re I_2$.


\begin{thebibliography}{999}
\bibitem{FOOTOM}
Note that in Eq.(\ref{d2}) we have omitted certain terms of order $t^{3/2}$,
e.g., the one proportional to $Re \la0|\h|\Psi^R\ra\Delta \Psi$. This term, however,
does not affect the asymptote of the current: it vanishes unless $\Delta \Psi\ne 0$,
but in this case we already have $P^R\sim \sqrt{t}$. For the same reason also unimportant is
the third term in the second curly bracket of Eq.(\ref{d2}).
\end{thebibliography}
\end{document}